\pdfoutput=1
\documentclass[aps,prl,reprint,superscriptaddress,longbibliography,amsmath,amssymb,floatfix]{revtex4-2}

\usepackage{graphicx}
\usepackage{bm}
\usepackage{xcolor}
\usepackage{multirow}
\usepackage{longtable}
\usepackage{tikz}
\usetikzlibrary{arrows.meta,positioning,calc}
\usepackage[colorlinks=true,linkcolor=blue,citecolor=blue,urlcolor=blue]{hyperref}

\newif\ifwithsm
\withsmtrue

\newcommand{\rth}{\rho_{\rm th}}
\newcommand{\rsc}{\rho_{\rm sc}}
\newcommand{\Occ}{\mathcal{O}}
\newcommand{\Tobs}{T_{\rm obs}}
\newcommand{\Lmax}{\mathcal{L}_{\max}}

\makeatletter
\def\author@fn#1{#1}
\def\author@fn@first#1{\frontmatter@footnote{#1}\let\author@fn\author@fn@next}
\def\author@fn@next#1{{\comma@space}\frontmatter@footnote{#1}}
\def\@author@join@#1#2#3{\def\@author{{#1}{#2\author@fn{#3}}}}
\def\author@nofn{\endgroup{}{}}
\def\author@withfn{\endgroup{\comma@space}{}\let\author@fn\author@fn@first\author@list}
\def\doauthor#1#2#3{%
  \ignorespaces#1\unskip\@listcomma
  \def\author@list{#2}%
  \begingroup
   #3%
  \ifx\author@list\@empty\expandafter\author@nofn\else\expandafter\author@withfn\fi
  \space \@listand
}
\makeatother

\begin{document}

\title{Toward Solving the Galactic Foreground Problem in Gravitational-Wave Astronomy with Machine Learning}

\affiliation{Center for Gravitational Wave Experiment, National Microgravity Laboratory, Institute of Mechanics, Chinese Academy of Sciences, Beijing 100190, China}
\author{Bo Liang}
\thanks{These authors contributed equally to this work.}
\affiliation{Center for Gravitational Wave Experiment, National Microgravity Laboratory, Institute of Mechanics, Chinese Academy of Sciences, Beijing 100190, China}
\author{Jinwen Deng}
\thanks{These authors contributed equally to this work.}
\affiliation{Kavli Institute for Astronomy and Astrophysics, Peking University, Beijing 100871, China}
\author{Minghui Du}
\email{duminghui@imech.ac.cn}
\affiliation{Center for Gravitational Wave Experiment, National Microgravity Laboratory, Institute of Mechanics, Chinese Academy of Sciences, Beijing 100190, China}
\author{Hanlin Song}
\affiliation{Center for Gravitational Wave Experiment, National Microgravity Laboratory, Institute of Mechanics, Chinese Academy of Sciences, Beijing 100190, China}
\author{Xilong Fan}
\email{xilong.fan@whu.edu.cn}
\affiliation{School of Physics and Technology, Wuhan University, Wuhan 430072, China}
\author{Peng Xu}
\email[Corresponding author: ]{xupeng@imech.ac.cn}
\affiliation{Center for Gravitational Wave Experiment, National Microgravity Laboratory, Institute of Mechanics, Chinese Academy of Sciences, Beijing 100190, China}
\author{Ziren Luo}
\affiliation{Center for Gravitational Wave Experiment, National Microgravity Laboratory, Institute of Mechanics, Chinese Academy of Sciences, Beijing 100190, China}
\author{Lijing Shao}
\affiliation{Kavli Institute for Astronomy and Astrophysics, Peking University, Beijing 100871, China}

\date{\today}

\begin{abstract}
Space-borne gravitational-wave detectors will see tens of millions of Galactic binaries. Resolved, they would form a census of the Galaxy's ultracompact binaries unhindered by dust; unresolved, they are the dominant noise up to a few millihertz for massive black-hole binaries, extreme-mass-ratio inspirals and backgrounds from the early Universe. Separating the two requires the number of binaries, which is ill-defined because binaries exist at every amplitude. We make it an observable by defining a source as a binary above a signal-to-noise threshold. Neural flows count the binaries to within one in 99\% of $1~\mu$Hz intervals over $0.5$--$10$~mHz, and nested sampling resolves them, even pairs closer than one frequency bin. The data themselves set a limit: a source is resolvable only if its signal pays for its own description, about 20~nats, which biases even ideal trans-dimensional samplers. Like the diffraction limit in imaging, this Bayesian confusion limit is set by the data rather than by the method, and it governs any superposition of weak sources, from radio surveys and pulsar timing arrays to super-resolution microscopy. For future space-borne detectors, the method turns the Galactic foreground into a definite, calibrated count of binaries with an analytic selection function, the basis for global fits and for tests of population models.
\end{abstract}

\maketitle

Counting and resolving signals that overlap in unknown numbers is a problem that recurs across physics and astronomy, from the confusion of faint radio sources~\cite{Scheuer1957,Condon1974} to the pile-up of collisions at particle colliders~\cite{Komiske2017} and crowded images in fluorescence microscopy~\cite{Holden2011}. One of its most extreme instances may be the millihertz sky of space-borne gravitational-wave (GW) observatories---LISA~\cite{AmaroSeoane2017,Armano2016,Colpi2024}, Taiji~\cite{Hu2017,Ruan2020} and TianQin~\cite{Luo2016}. Unlike the sparse transient sky of ground-based detectors~\cite{Abbott2016,GWTC3}, it is filled by tens of millions of continuously radiating compact binaries in the Milky Way, mostly double white dwarfs~\cite{Nelemans2001,Korol2017,Hermes2012,Burdge2019}. Of order $10^4$ of them will be resolvable; the rest blend into a stochastic foreground that exceeds the instrument noise up to a few mHz~\cite{Timpano2006,Robson2019,Karnesis2021}. This foreground is at once a treasure and an obstacle. Resolved into binaries, it provides a census of the Galaxy's ultracompact binaries that no electromagnetic telescope can take, because GWs propagate through the dust of the Galactic plane, with one to two orders of magnitude more systems than optical surveys will find~\cite{Korol2017}; it probes binary evolution~\cite{Breivik2018}, tides and general relativity in compact orbits~\cite{Burdge2019,Fiacco2024}, the progenitors of Type~Ia supernovae~\cite{Webbink1984}, and the structure of the Milky Way and its satellites~\cite{Korol2017,Roebber2020}. Left unresolved, it is the dominant noise up to a few mHz for every other source: massive black-hole binaries that trace the growth of galaxies, extreme-mass-ratio inspirals that test strong gravity and the nature of dark matter~\cite{Cardoso2022,Speri2023,Duque2024}, and backgrounds from inflation or cosmological phase transitions~\cite{Adams2014,Hindmarsh2014,Littenberg2023}.

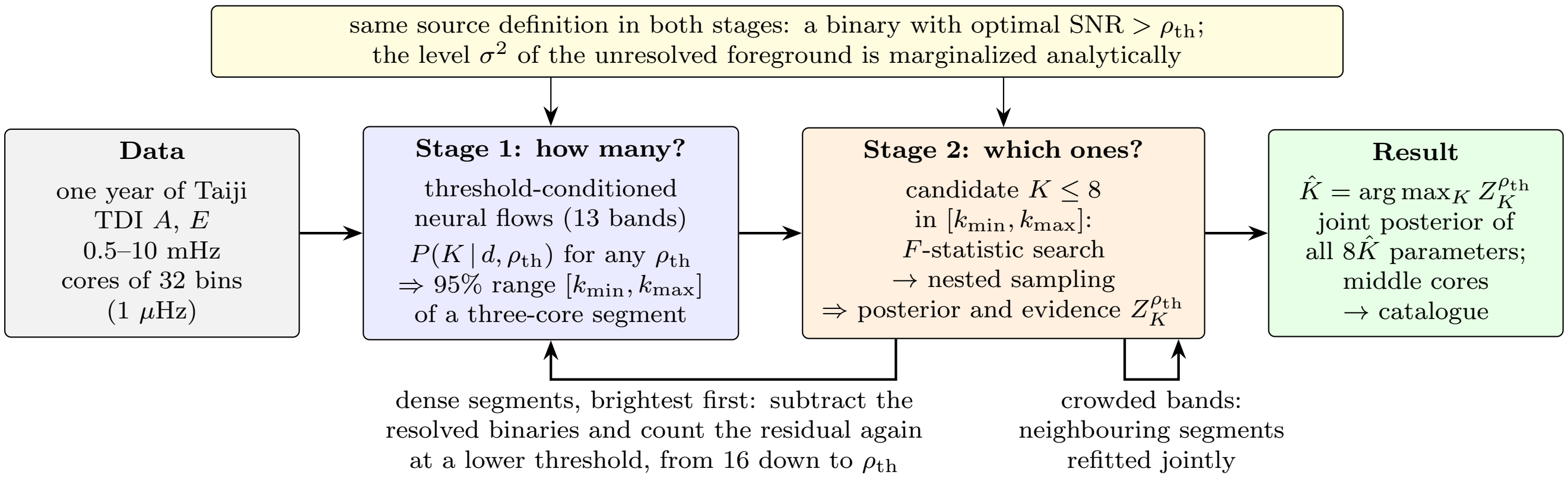
\begin{figure*}[t]
\centering
\begin{tikzpicture}[font=\footnotesize, >={Stealth[length=2.2mm]},
  box/.style={draw, rounded corners=2pt, align=center, inner sep=4pt, minimum height=2.15cm, text width=#1}]
\thickmuskip=5mu\relax\medmuskip=4mu\relax
\node[box=3.0cm, fill=gray!10] (data) {\textbf{Data}\\[3pt] one year of Taiji\\ TDI $A$, $E$\\ $0.5$--$10$~mHz\\ cores of 32 bins\\ ($1~\mu$Hz)};
\node[box=3.9cm, fill=blue!8, right=0.7cm of data] (s1) {\textbf{Stage 1: how many?}\\[3pt] threshold-conditioned\\ neural flows (13 bands)\\[2pt] $P(K\,|\,d,\rth)$ for any $\rth$\\ $\Rightarrow$ 95\% range $[k_{\min},k_{\max}]$\\ of a three-core segment};
\node[box=4.2cm, fill=orange!12, right=0.7cm of s1] (s2) {\textbf{Stage 2: which ones?}\\[3pt] candidate $K\le8$ in $[k_{\min},k_{\max}]$:\\ $F$-statistic search\\ $\to$ nested sampling\\ $\Rightarrow$ posterior and evidence $Z_K^{\rth}$};
\node[box=3.0cm, fill=green!10, right=0.7cm of s2] (out) {\textbf{Result}\\[3pt] $\hat K=\arg\max_K Z_K^{\rth}$\\ joint posterior of\\ all $8\hat K$ parameters;\\ middle cores $\to$ catalogue};
\draw[->, thick] (data) -- (s1);
\draw[->, thick] (s1) -- (s2);
\draw[->, thick] (s2) -- (out);
\node[draw, rounded corners=2pt, fill=yellow!15, align=center, inner sep=3pt, text width=12.4cm] (def) at ($(s1.north)!0.5!(s2.north)+(0,0.95)$)
  {same source definition in both stages: a binary with optimal $\mathrm{SNR}>\rth$;\\ the level $\sigma^2$ of the unresolved foreground is marginalized analytically};
\draw[->] (def.south -| s1.north) -- (s1.north);
\draw[->] (def.south -| s2.north) -- (s2.north);
\draw[->, thick] ($(s2.south)+(-1.2,0)$) -- ++(0,-0.45) -| node[pos=0.35, below, align=center, text width=7.0cm]
  {dense segments, brightest first: subtract the\\ resolved binaries and count the residual again\\ at a lower threshold, from 16 down to $\rth$} (s1.south);
\draw[->, thick] ($(s2.south)+(1.35,0)$) -- ++(0,-0.45) -| node[pos=0.25, below, align=center, text width=3.4cm]
  {crowded bands:\\ neighbouring segments\\ refitted jointly} ($(s2.south)+(1.95,0)$);
\end{tikzpicture}
\caption{The analysis architecture. Stage~1 (machine learning) returns, for any threshold $\rth$, the posterior of the number $K$ of binaries in each 32-bin core; Stage~2 (Bayesian inference) locates and samples the binaries of each candidate $K$, and the evidence $Z_K^{\rth}$ selects $\hat K$. In dense segments the two stages alternate, brightest first: the binaries counted above a high threshold are resolved and subtracted, and the residual is counted again at a lower one; in the densest segments of the crowded bands, neighbouring segments are refitted jointly (see Supplemental Material).}
\label{fig:flowchart}
\end{figure*}

\begin{figure*}[t]
\includegraphics[width=\textwidth]{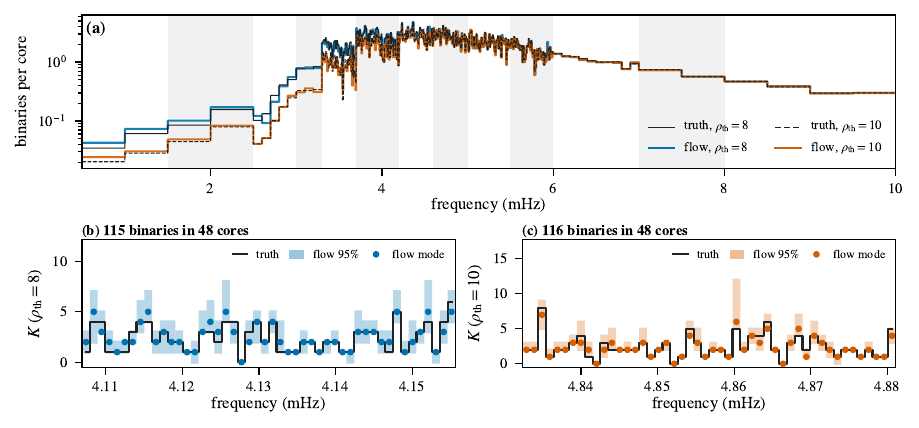}
\caption{Binaries counted by 13 band flows in 9330 cores of 32 bins ($1~\mu$Hz) over $0.5$--$10$~mHz. (a) Binaries per core with SNR above $\rth=8$ and $10$ in cells of $0.01$--$0.5$~mHz: truth (black) and flows (colored); shading alternates between flows. (b),(c) Truth (line), 95\% interval (shading) and most probable count (dots) per core at $\rth=8$ and $10$, for the densest stretch of 48 consecutive cores in which every count is within $\pm1$ of the truth and every truth inside the interval.}
\label{fig:count}
\end{figure*}

Separating this foreground is hard for two reasons, one practical and one conceptual. Its signals overlap~\cite{Littenberg2020}: the annual motion of the detector spreads each binary over $\pm10$ frequency bins at 3~mHz~\cite{Cutler1998,CornishLarson2003}, and in the $3.0$--$3.3$~mHz band of our simulation half of the 384 binaries above a signal-to-noise ratio (SNR) of 7 have a neighbour within this range, 22 within a single bin. And their number is not even well defined: binaries exist at every amplitude, down to the millions that make up the foreground~\cite{Nelemans2001,Timpano2006}, so, as for radio sources below the confusion limit~\cite{Scheuer1957,Condon1974}, a count depends on where one stops counting, that is, on how one defines signals too faint to detect~\cite{Littenberg2020}. Trans-dimensional samplers, the standard tool, treat the number of binaries $K$ as a free parameter~\cite{Green1995,Crowder2007,Littenberg2011,Littenberg2020,Katz2024}; they inherit this ambiguity and, exploring a space whose dimension grows as $8K$, mix and converge slowly in crowded data~\cite{Littenberg2020,Korsakova2024}. Global fits built on them remain the reference on the simulated data of the LISA Data Challenges~\cite{Littenberg2023,Lackeos2023,Strub2024,Katz2024}, but the binaries dominate their cost---about a week on four GPUs~\cite{Katz2024}---and their catalogues inherit the prior dependence. Machine learning has made the inference of isolated signals fast~\cite{Chua2020,Dax2021,Gabbard2022,Dax2025,Du2024,Liang2024} and has begun to speed up global fits~\cite{Korsakova2024} and to characterize pairs of overlapping signals~\cite{Langendorff2023}. Yet the number of sources---the trans-dimensional core of the problem---has so far been left to samplers, and what is being counted, to the prior.

We instead treat the number of sources as a physical observable. With the definition used for radio sources below the confusion limit~\cite{Scheuer1957,Condon1974} and in forecasts of the resolvable population~\cite{Korol2017}, a source is a binary whose optimal SNR exceeds a threshold $\rth$. Fainter binaries then belong to the foreground rather than to the model, so the prior no longer has to weigh signals too faint to detect: the number of sources $K(\rth)$ becomes an observable, and its Bayesian evidence, the probability $Z_K=p(d\,|\,K)$ of the data given $K$ binaries averaged over their prior, is well defined: for four injected binaries, a fifth is rejected by $\ln B_{4,5}=\ln(Z_4/Z_5)=+10.7$, whereas without the threshold a scale-free amplitude prior leaves it a probability of 0.28 (see Supplemental Material). Measuring this observable and resolving its sources are different questions, and we hand each to the tool suited to it (Fig.~\ref{fig:flowchart}). The discrete question, how many binaries, goes to machine learning (Stage~1): threshold-conditioned neural flows~\cite{Gulati2020,Durkan2019,Papamakarios2021}, trained on simulations alone~\cite{Cranmer2020}, return the posterior of $K$ in each frequency core for every $\rth$ in a single evaluation. The subsequent question, which binaries, goes to exact Bayesian inference (Stage~2): for each $K$ in the flow's 95\% interval, an $F$-statistic~\cite{Jaranowski1998} matching pursuit~\cite{Mallat1993,CornishLarson2003} locates the binaries and nested sampling~\cite{Skilling2006,Williams2021} draws their joint posterior from the full likelihood, in which the unresolved foreground is an unknown noise scale $\sigma^2$ marginalized with a Jeffreys prior~\cite{Jeffreys1961,Rover2011}; the prior excludes binaries below $\rth$, so the evidence refers to the same sources as the flow (see Supplemental Material). Machine learning thus removes the combinatorial bottleneck, while every reported binary keeps an exact posterior and evidence. We test this on one year of simulated data of the Taiji constellation in the time-delay-interferometry channels $A$ and $E$~\cite{Tinto2021}, divided into cores of 32 frequency bins ($1~\mu$Hz), the unit in which the flows count. Because signals leak into neighbouring cores, Stage~2 fits a segment of three adjacent cores but keeps in the catalogue only the binaries of the middle core; with a segment centred on every core, every binary is fitted together with its neighbours and catalogued once. Binaries above SNR 7 are drawn from the Galactic-binary catalogue of the Taiji Data Challenge~II~\cite{Du2025} and injected individually, while the fainter ones enter the noise spectrum as confusion noise.

\begin{table}[tb]
\caption{Counting on 40 random windows of $3.0$--$3.3$~mHz: cores (\%) counted exactly by an ideal trans-dimensional sampler, which takes the $K$ of largest evidence (``evidence''), and by the flow, and mean count bias per core, for the evidence measured and predicted. Standard errors over the windows: 4--8 points for the percentages (see Supplemental Material).}
\label{tab:confusion}
\resizebox{\columnwidth}{!}{%
\begin{tabular}{lccccc}
\hline\hline
 & \multicolumn{2}{c}{exact (\%)} & \multicolumn{3}{c}{bias per core} \\
$\rth$ & evidence & flow & evidence & predicted & flow \\
\hline
8 & 77.5 & 80.0 & $+0.28\pm0.11$ & $+0.10$ & $+0.03\pm0.08$ \\
9 & 67.5 & 77.5 & $+0.48\pm0.13$ & $+0.34$ & $-0.08\pm0.08$ \\
10 & 67.5 & 92.5 & $+0.45\pm0.12$ & $+0.44$ & $-0.08\pm0.04$ \\
12 & 62.5 & 95.0 & $+0.45\pm0.10$ & $+0.49$ & $0.00\pm0.04$ \\
\hline\hline
\end{tabular}}
\end{table}

The division is not merely convenient: the data themselves limit what the evidence can count. Adding a binary to a model changes the log evidence by
\begin{equation}
\Delta\ln Z=\Delta\ln\Lmax-\Occ,
\label{eq:occam}
\end{equation}
where the Occam factor $\Occ$ is the information the data provide about the binary~\cite{MacKay2003}; over the 1461 binaries added in our Stage-2 runs, $\Occ\approx20$~nats, most of it from the sky position and the frequency. Since $\Delta\ln\Lmax\simeq\hat\rho^{\,2}/2\sigma^2$ for a binary of measured SNR $\hat\rho$ in a foreground of level $\sigma$, a binary is resolvable only if its signal pays for its own description, $\hat\rho^{\,2}/2\sigma^2>\Occ$, a minimum-description-length criterion~\cite{Rissanen1978} for GW sources. This Bayesian confusion limit, the counterpart of the confusion limit of radio surveys~\cite{Condon1974}, $\rsc=\sqrt{2\Occ}\,\sigma\approx7$, does not move with the threshold one chooses. Above it, a binary just below $\rth$ can still be fitted at the prior boundary, losing only $(\rth-\hat\rho)^2/2\sigma^2$ of likelihood, so the evidence admits every binary of measured SNR above
\begin{equation}
\rho^*=\frac{\rth^2+\rsc^2}{2\rth}\qquad(\rth\ge\rsc),
\label{eq:rhostar}
\end{equation}
$8.0$ at $\rth=10$ for $\rsc=7.8$, the value measured on our earlier windows, and counts real binaries between $\rho^*$ and $\rth$; below $\rsc$, binaries between $\rth$ and $\rsc$ cannot pay and are missed. The limit is self-consistent: resolving more binaries lowers the foreground and with it $\rsc$~\cite{Timpano2006}, so the boundary between resolved binaries and the foreground is a fixed point set by the data, at the SNR${}\approx7$ adopted by convention in forecasts~\cite{Korol2017}.

An ideally converged trans-dimensional sampler draws $K$ with probability $\propto P(K)\,Z_K$~\cite{Green1995} and so inherits the selection of Eq.~(\ref{eq:rhostar}): it detects each binary with probability $\Phi[(\rho-\rho^*)/\sigma]$, with $\Phi$ the standard normal distribution function. With $\Occ$ and $\sigma$ measured on earlier windows, we predicted the count bias at every threshold before measuring it. On 40 random windows of $3.0$--$3.3$~mHz, taking the $K$ of largest evidence returns the true count in 63--78\% of the cores at $\rth=8$--$12$, and its bias follows the prediction (Table~\ref{tab:confusion}), crossing zero at $\rth=7.3$, where the theory places $\rsc$ ($7.2$--$7.8$ for the foreground levels we measure); derived from first principles in the Supplemental Material, the same theory predicts the fraction of cores counted exactly within $2.3$ points at $\rth=8$--$12$. The flow, in contrast, learns the chosen definition from its labels, including how often fainter binaries scatter above the threshold, stays unbiased and returns the true count in 78--95\% of the cores. Counting and resolving remain different questions, however: imposing the flow's count on the sampler resolves fewer faint binaries (see Supplemental Material), so the flow sets the range of $K$ and the evidence chooses within it.

Across the whole band the flows determine $K$ to within one binary almost everywhere, and where faint binaries dominate they reach the limit set by noise. Thirteen flows, one trained for each frequency band, count the binaries in all 9330 cores between $0.5$ and $10$~mHz (Fig.~\ref{fig:count}; per band in Table~D1 of the data release; see Supplemental Material). The count is within one of the truth in $99.0\%$ of the cores at $\rth=8$ and $99.3\%$ at $\rth=10$, and the flow's 95\% interval contains the truth in $99\%$. No method can count every core exactly, because the measured SNR of a binary scatters by about one around its true value: an oracle that knows every injected waveform and is ignorant only of the noise counts at most $90.7\%$ of the cores of $3.0$--$3.3$~mHz exactly at $\rth=10$, and the flow reaches $90.8\%$. Below $3.3$~mHz every flow comes within $1.1$ points of this information limit at $\rth=10$ ($5.6$ at $\rth=8$), so counting noise at the threshold, the GW analog of Eddington bias~\cite{Eddington1913}, is its only significant error. Where up to 14 bright binaries share a core, the flows undercount but their intervals still bracket the truth, so the sampler searches the right $K$.

Counting is also a measurement of the Galaxy. Because one evaluation covers every threshold, the flow measures the source counts $N(>\rho)$, the statistic with which radio astronomers first probed the distribution of radio sources---and which confusion famously distorted~\cite{Ryle1955,Scheuer1957,DeZotti2010}. In $3.0$--$3.3$~mHz the flow's expected counts follow $N(>\rho)\propto\rho^{-\gamma}$ with $\gamma=3.39\pm0.22$, against the true $3.34\pm0.23$, free of confusion bias; a uniform population in Euclidean space would give $\gamma=3$ for any luminosity function, so the slope measures how the binaries are distributed in the Galaxy.

\begin{figure*}[t]
\includegraphics[width=\textwidth]{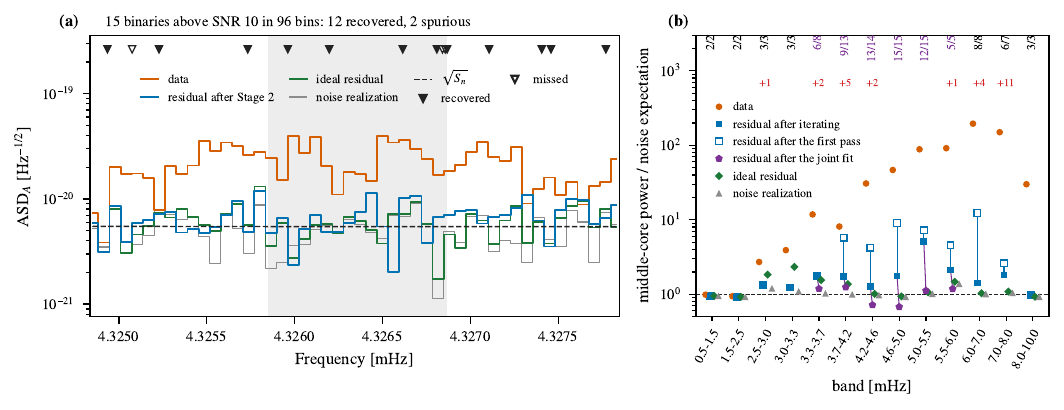}
\caption{Resolving the densest segments, from the brightest binaries down. (a) Amplitude spectral density of channel $A$, averaged over pairs of bins, over a segment of $4.2$--$4.6$~mHz (middle core shaded): data, residual after Stage~2, ideal residual (data minus every injected binary above SNR~10), noise realization and $\sqrt{S_n}$; triangles mark the binaries above SNR~10, filled when recovered. (b) Middle-core power of the segment with the highest flow count at $\rth=10$ in each of the 13 bands, relative to the noise expectation, with the residual after the first pass (open squares), iterating (filled) and fitting neighbouring segments jointly (purple); on top, recovered over true binaries above SNR~10 and spurious components (red) of the last fit. Where the joint-fit residual falls below the noise realization, it stays above the floor of a correct fit, which absorbs the noise its waveforms can describe (see Supplemental Material).}
\label{fig:densemain}
\end{figure*}

Resolution, too, is set by physics rather than by the spectrum. Two binaries closer than the frequency resolution $1/\Tobs$ have indistinguishable spectra, but the annual orbit of the constellation modulates the phase and amplitude of each signal according to its sky position, so the detector acts as a synthetic aperture two astronomical units across~\cite{Cutler1998,CornishLarson2003}. We test this on six windows, each with three to five injected binaries whose signals overlap in the spectrum, adjacent ones as close as a third of a frequency bin and four within 1.4 bins in one of them (see Supplemental Material). The evidence selects the true number every time, the recovered waveforms match the injected ones to $0.96$--$0.997$, and 198 of the 208 injected parameters ($95.2\%$) lie inside the 95\% credible intervals. Frequencies are measured to ${\approx}\,2$~nHz, $1/14$ of a bin, and positions to $4$--$56$~deg$^2$.

On a realistic band the pipeline yields a catalogue: on the 293 segments of $3.0$--$3.3$~mHz at $\rth=10$, with $K$ restricted to the flow's range, it returns 88 confident detections (posterior median SNR $\ge\rth+1$), $97.7\%$ of them physical binaries, whose number per core is exact in $91.8\%$ of the cores, matching the information limit of $90.7\%$. It recovers all 55 binaries with SNR above 12 and separates pairs as close as $0.14$ frequency bins, and subtracting it leaves only noise and the binaries below the source definition (data release; see Supplemental Material). Such a catalogue reaches the Galactic plane, where 89\% of our resolvable binaries lie and where extinction hides double white dwarfs from optical surveys; for the $\sim10^4$ resolvable binaries of a full analysis, it would exceed the number of double white dwarfs expected from Gaia and the Vera C. Rubin Observatory by one to two orders of magnitude~\cite{Korol2017}.

The crowded bands of $3.3$--$6.0$~mHz, where binaries are densest in frequency, test the method where confusion is strongest. Counting first takes Stage~2 past the cap of eight binaries that we place on one nested-sampling run: in denser segments it resolves the binaries counted above SNR~16, subtracts them and repeats on the residual at lower thresholds. In a segment of $4.2$--$4.6$~mHz with 15 injected binaries above SNR~10 in 96 bins, this brightest-first subtraction recovers 12 of them, with two spurious components (fits that match no injected binary), and lowers the power in the middle core, the part kept in the catalogue, from 28 times the noise expectation to $1.35$, against $1.13$ after subtracting the injected binaries exactly [Fig.~\ref{fig:densemain}(a)]; in the other three segments of the band counted at 16 it recovers 5 to 8 of their 16. On the segment with the highest flow count in every band [Fig.~\ref{fig:densemain}(b)], iterating lowers the middle-core residual up to ninefold; every binary is recovered where at most three exceed SNR~10, but only 61\% in the crowded bands of $3.3$--$6.0$~mHz; fitting neighbouring segments jointly, each refitted in turn with the binaries of the others subtracted, raises this to 86\% and cuts spurious components from 41 to 10 (see Supplemental Material). A single pass recovers every binary in the five sparse bands ($0.5$--$3.3$ and $8$--$10$~mHz), leaving $0.92$--$1.35$ times the noise expectation; iterating leaves $1.41$--$1.82$ at $6$--$8$~mHz, and the joint fit $0.68$--$1.25$ at $3.3$--$6.0$~mHz, at or below the ideal residual in five of the six bands. One pass thus suffices where binaries are sparse, and a joint fit is needed where they crowd.

\textit{Outlook.}---Treating the foreground as a counting problem first and a parameter-estimation problem second changes what space-borne GW astronomy can do with it. With unbiased counts and the analytic selection function of Eq.~(\ref{eq:rhostar}), the census can directly test population-synthesis forecasts~\cite{Korol2017} and enter population inference without the bias of an unmodelled detection efficiency~\cite{Mandel2019}, so that the counts and the slope of the source counts become direct probes of binary evolution, of the progenitors of Type~Ia supernovae and of the structure of the Galaxy~\cite{Webbink1984,Korol2017,Breivik2018}. Resolving close pairs with calibrated sky positions in every band could map the binaries' distribution and substructure across the Galaxy~\cite{Korol2017,Roebber2020}, a first step that longer data will refine. Determined by the data rather than by convention, the unresolved remainder becomes a calibrated noise against which massive black-hole binaries that trace the growth of galaxies, extreme-mass-ratio inspirals that test strong gravity and the nature of dark matter~\cite{Cardoso2022,Speri2023,Duque2024}, and primordial backgrounds from inflation or phase transitions~\cite{Adams2014,Hindmarsh2014} can be sought, while the counts and posteriors supply the source numbers and starting points that global fits require~\cite{Littenberg2023,Katz2024}, pointing to a machine-learning-based global fit that pairs amortized counting with exact sampling. Because the flows are amortized and every window is analyzed independently~\cite{Cranmer2020}, the method extends to the full band and to multi-year data at the cost of computing time alone. Two steps remain before real data: because the flows learn the population they are trained on (see Supplemental Material), the population and the noise spectrum must be fitted iteratively together with the catalogue; and non-stationary noise, data gaps and the other millihertz sources must enter the simulated windows. More broadly, the principle we identify---a source is resolvable only if its signal pays for its own description, ${\approx}\,20$~nats here---sets a limit of the data rather than of the method, as the diffraction limit does for imaging, and governs any superposition of weak sources: confusion in radio surveys~\cite{Condon1974} and in infrared and X-ray surveys; overlapping signals in third-generation ground-based detectors~\cite{Punturo2010,Reitze2019,Samajdar2021,Langendorff2023} and the GW background of pulsar timing arrays~\cite{Agazie2023,Xu2023}; the number of planets in radial-velocity data~\cite{BrewerDonovan2015}; pile-up of collisions at particle colliders~\cite{Komiske2017}; and overlapping emitters in super-resolution microscopy~\cite{Cox2012}, some of whose algorithms descend from crowded-field astronomy~\cite{Holden2011}. Learning how many sources there are, and inferring which ones they are, offers a template for such problems.

\begin{acknowledgments}
We thank the Taiji Data Challenge team for the Triangle simulator, Huan Yang (Tsinghua University) for discussions, and Alvin Chua for the EMRI workshop he organized at the National University of Singapore. This work is supported by the National Key R\&D Program of China under Grants No.~2025YFE0217300 and the International Partnership Program of the Chinese Academy of Sciences (Grant No.~025GJHZ2023106GC). JD was supported by the ``AI+X'' program for the Undergraduate Student Research Study at Peking University. LS was supported by the Beijing Natural Science Foundation (1242018).
\end{acknowledgments}

\bibliography{references}

\ifwithsm
\clearpage
\section*{Supplemental Material}
\setcounter{figure}{0}
\setcounter{table}{0}
\setcounter{equation}{0}
\renewcommand{\thefigure}{S\arabic{figure}}
\renewcommand{\theHfigure}{S\arabic{figure}}
\renewcommand{\thetable}{S\arabic{table}}
\renewcommand{\theHtable}{S\arabic{table}}
\renewcommand{\theequation}{S\arabic{equation}}
\renewcommand{\theHequation}{S\arabic{equation}}
\makeatletter\setlength{\@fptop}{0pt}\makeatother
\subsection*{Simulated data}

\textit{Detector and noise.}---We simulate one year ($\Tobs=3.15\times10^7$~s, sampled every 10~s) of a Taiji-like constellation with Triangle, the simulator of the Taiji Data Challenge~II~\cite{Du2025}, and analyze the time-delay-interferometry channels $A$ and $E$~\cite{Tinto2021} in the frequency domain, in bins of width $1/\Tobs=31.7$~nHz. The noise is Gaussian with a reference spectrum $S_n(f)$ that contains the instrument noise and the confusion noise of the fainter Galactic binaries. Data are whitened by dividing every bin by $(S_n\Tobs/4)^{1/2}$, so that the real and imaginary parts of the noise have unit variance.

\textit{Galactic binaries.}---Binaries are drawn from the Galactic-binary catalogue of the Taiji Data Challenge~II~\cite{Du2025}. Their frequencies follow the catalogue's source density, histogrammed in bins of 1~$\mu$Hz; their optimal SNRs are drawn from the SNR distribution of the catalogue binaries in the same band, capped at 100 (median 8--11 below 4.2~mHz, rising to 96 at 8--10~mHz), and their frequency derivatives $\dot f_0$ and sky positions jointly from its binaries; inclination, initial phase and polarization angle are isotropic. Each waveform is computed with Triangle's Galactic-binary generator, combined into $A$ and $E$, and rescaled so that its optimal SNR against $S_n(f)$ equals the drawn value. Only binaries with $\mathrm{SNR}\ge7$ are injected, because $S_n(f)$ already contains the confusion noise of the fainter ones.

\textit{Cores and segments.}---The data are divided into cores of 32 bins ($1~\mu$Hz). Stage~1 reads each core together with 48 bins of context on each side, 128 bins in total. Stage~2 analyzes segments of three adjacent cores, a 96-bin prior region, with 16 further bins of data on each side; it keeps only the binaries recovered in the middle core, so the outer cores absorb leakage from bright neighbours, and one segment is centred on every core.

\textit{Analysis and test bands.}---The joint analysis covers the 293 cores of $3.0$--$3.3$~mHz. This band holds 384 binaries with $\mathrm{SNR}\ge7$: half of them have another within $\pm10$ bins, the range over which the annual Doppler modulation spreads a signal at 3~mHz, 22 have one within a single bin, and 89\% lie within $5^\circ$ of the Galactic plane. Each of the 13 band flows is tested on a continuous realization of its band, drawn from the catalogue as above with an independent random seed and never used in training or validation, without the outermost core on each side; the $3.0$--$3.3$~mHz flow is tested on the analysis band. Together the test bands hold 9330 cores.

\subsection*{Source definition and evidence}

\textit{Model selection without a threshold.}---For a benchmark injection of four overlapping binaries, a log-uniform amplitude prior prefers $K=4$ over $K=5$ by only $\ln B_{4,5}=3.7$ inside the localization box and by at most $0.94$ once the prior volume of an invisible fifth binary outside the box is included, i.e.\ $P(K=5)\approx0.28$. Restricting every binary to $\mathrm{SNR}>8$ gives $\ln B_{4,5}=+10.7$, the fifth binary fitting $\mathrm{SNR}=4.4\,[3.3,5.5]$, and $\ln B_{4,3}=+31.8$. Fig.~D1 of the data release shows how the threshold restores model selection: without it (triangle), the evidence of $K=5$ lies within one nat of that of the true $K=4$; requiring every binary to exceed $\rth$ lowers it steadily, to $-10.7$ at $\rth=8$, while $K=3$ is rejected by about 32~nats at every threshold. Its panel~(b) shows what the fifth component fits: a binary of $\mathrm{SNR}\approx4$, which the data cannot rule out unless the prior excludes it.

\textit{Likelihood.}---The unresolved foreground enters as an unknown scale of the noise spectrum, $\sigma^2 S_n(f)$, with a Jeffreys prior $\pi(\sigma^2)\propto\sigma^{-2}$~\cite{Jeffreys1961}; marginalizing $\sigma^2$ gives
\begin{equation}
\ln\mathcal{L}(\bm\theta)=\ln\Gamma(N)-N\ln\left[Q(\bm\theta)/2\right],
\label{eq:likelihood}
\end{equation}
where $Q$ is the noise-weighted residual power of $K$ binary waveforms over the $N$ complex data points of both channels. This Gaussian likelihood integrated over the noise level behaves like a Student-$t$ distribution in the residuals~\cite{Rover2011} and adapts to the local foreground; its maximum gives the foreground level of a segment, $\hat\sigma^2=Q/2N$. Each binary has eight parameters $(A,f_0,\dot f_0,\iota,\phi_0,\lambda,\beta,\psi)$ with priors log-uniform in $A$ over $(10^{-24},10^{-20.5})$, uniform in $f_0$ over the prior region and in $\dot f_0$ over $(-5\times10^{-17},3\times10^{-16})$~Hz/s, and isotropic in sky position and orientation; the $K$ binaries are ordered in $f_0$, which removes the $K!$ relabelings of the same solution.

\textit{Threshold-consistent evidence.}---The threshold-truncated prior multiplies the prior density by $\Theta[\mathrm{SNR}(\bm\theta)-\rth]/p_\rho$, so the evidence of $K$ binaries is
\begin{equation}
\ln Z_K^{\rth}=\ln Z_{\rm box}+\ln V_{\rm box}+\ln q_{\rm box}-K\ln p_{\rho},
\label{eq:evidence}
\end{equation}
where $Z_{\rm box}$ is the nested-sampling evidence inside a localization box, $V_{\rm box}$ the prior mass of that box under the ordered prior, $q_{\rm box}$ the fraction of the box in which every binary exceeds $\rth$, and $p_\rho$ the prior probability that one binary does. $V_{\rm box}$, $q_{\rm box}$ and $p_\rho$ are estimated by Monte Carlo, and their binomial errors are propagated. Fixing $\sigma=1$ instead of marginalizing it raises the evidence-alone count bias at $\rth=10$ from $+0.45$ to $+0.70\pm0.15$ binaries per core.

\subsection*{Stage 1: count flows}

\textit{Network.}---The input of a core is the whitened data of the core and its 48 context bins on each side, 128 bins of four channels: the real and imaginary parts of $A$ and $E$. A convolution over five neighbouring bins, followed by a SiLU activation, maps the input to one token per bin, of the width of the network, to which learned position and region embeddings, marking core and context, and an embedding of the core's frequency within the band are added. Six Conformer blocks~\cite{Gulati2020} (width 144 with four attention heads, or 256 with eight; convolution kernel 7, dropout $0.1$) process the 128 tokens. This encoder never sees the threshold. The threshold enters only the count head: a small network embeds the log-scaled threshold, and the 32 core tokens are pooled once per threshold by their mean, by softmax attention and by a sigmoid token gate, the last two conditioned on the threshold embedding, so that the head can decide which local peaks count at a given threshold. The count distribution is a monotone flow on the bounded count coordinate $[-0.5,K_{\max}+0.5]$~\cite{Durkan2019,Papamakarios2021}: a logistic base distribution transformed by four monotone layers gives the cumulative distribution at the edges between consecutive counts, whose differences are the probabilities of $K=0,\dots,K_{\max}$, and no probability is piled up at the ends of the range. One pass of the encoder therefore yields $P(K\,|\,d,\rth)$ at every threshold. An auxiliary head, used only as a training signal, decodes the tokens into a map of where the binaries above the threshold lie in the core, a Gaussian of width $0.3$ bins per binary on a grid eight times finer than the bins.

\textit{Training windows.}---Training windows are simulated on the GPU during training, so that no window is used twice. Each window is centred on a random core of the band and carries six thresholds drawn log-uniformly in $[7,16]$; its labels are the numbers of core binaries above each of them, and the first threshold builds the scene. A quarter of the windows keep the natural draw described above. So that every count up to $K_{\max}$ is trained, the other three quarters replace the binaries above the scene threshold by a number drawn uniformly from $0$ to $K_{\max}$, placed uniformly in frequency, with SNRs drawn from $p(\rho)\propto\rho^{-4.5}$ between the threshold and 100. In half of all windows this is done in the core and in every neighbouring core, whose counts are drawn uniformly from $0$--$2$, from $0$--$K_{\max}$ or from $K_{\max}/2$--$K_{\max}$; in a quarter it is done in the core only, and in 35\% of these windows with two or more binaries some of them are clustered around a common frequency, with a spread of $0.25$--$2$ bins. Binaries below the scene threshold keep their natural draw. The density of natural windows must match the data: power-law SNR labels ($\mathrm{SNR}_{\min}=7$, index 4) for every catalogue binary place nine to fourteen times more binaries above SNR 8 and 15 in each core than the catalogue, and a network trained with the binaries below SNR 7 injected as well, on top of the confusion noise already in $S_n(f)$, under-counts the band ($0.49$ instead of $0.79$ binaries per core at $\rth=8$) and is overconfident (its 95\% intervals cover $92.8\%$).

\textit{Optimization.}---The loss is the negative log-likelihood of the six labels, plus $0.15$ times the mean squared difference between the predicted and true cumulative count distributions, which penalizes probability far from the truth, plus $0.1$ times the squared error of the auxiliary map and $0.1$ times a smooth-$L_1$ penalty between its integral and the true count. We use AdamW with learning rate $3\times10^{-4}$ and weight decay $0.01$, a linear warm-up over 2000 steps followed by cosine decay to $10^{-6}$, gradient clipping at norm 1, mixed precision and an exponential moving average of the weights (decay $0.9995$), for 48 epochs of 1000 steps with 64 windows per GPU and step. Validation uses 2048 windows drawn like the training windows, 550 with balanced counts and 550 with dense neighbours, at $\rth=8$, $9$ and $12$. Every flow is tested with its final weights; the number of epochs, 24 or 48, is chosen on the validation accuracy at $\rth=8$.

\textit{Calibration.}---On held-out windows with 300 per true count from 0 to 6, the 68\% intervals of the $3.0$--$3.3$~mHz flow contain the truth in 73--98\% of the windows for every count when the neighbours are dense ($\rth=8$, $9$ and $12$); with natural neighbours they do so in 65--99\% up to four binaries per core and in 55--69\% for five or six, which are rare in this band. The mean width of the 68\% intervals is at most $1.3$. On the analysis band itself, the 68\% intervals cover the truth in 98\%, 89\%, 96\% and 98\% of the cores at $\rth=7$, $8$, $12$ and $16$.

\textit{Candidate range.}---The candidate range $[k_{\min},k_{\max}]$ of a Stage-2 segment is the 95\% shortest interval of the convolved count posteriors of its three cores, widened by one; it contains the true count in all 293 segments (290 without the widening).

\textit{Band flows.}---Each of the 13 flows is trained from scratch on simulated windows of its own band, in the same way as the $3.0$--$3.3$~mHz flow; besides the epochs and width chosen below, only the band and the capacity $K_{\max}$, 6 to 14 binaries per core, differ. Bands are $0.3$--$0.5$~mHz wide where binaries are dense and up to 2~mHz where they are sparse, because the density of binaries above $\rth=8$ ranges from $0.05$ to $3.2$ per core and the Doppler spread of a signal grows from $\pm3$ bins at 1~mHz to $\pm30$ at 10~mHz. The flows above $3.3$~mHz are trained for 48 epochs, which raises the validation accuracy over 24 epochs in every band; below $3.0$~mHz, where the flows are already within 3 points of the information limit, 48 epochs do not, and the 24-epoch flows are used. Between $3.7$ and $8$~mHz the flows have width 256, which validates better than width 144 in each of these seven bands, by $1.6$--$4.7$ points; in $3.3$--$3.7$~mHz the two widths validate equally and width 144 is kept, as in all other bands. The width is likewise chosen by validation accuracy. Four seeds of the width-144 flows scatter by $0.1$--$2.5$ points (standard deviation) in the fraction of cores counted exactly at $\rth=8$; in each of the seven bands width 256 gains more than this scatter, $1.0$--$3.6$ points averaged over two seeds, and the coverage of the 95\% intervals is unchanged. The flow of each band was trained in one to seven candidate versions, 65 in all: one for $3.0$--$3.3$~mHz, two for each of the four sparse bands (24 and 48 epochs) and seven for each of the eight bands of $3.3$--$8$~mHz (24 and 48 epochs at width 144, three more seeds at 48 epochs and width 144, and two seeds at width 256). One version per band, 13 in all, is used, selected as described above; the others served for this selection and to measure the scatter between seeds. Above $5.5$~mHz all but three catalogue binaries have $\mathrm{SNR}>10$ (the faintest $8.5$; all $>24$ above 8~mHz), so the counts there hardly depend on $\rth$. The totals of Table~D1 of the data release sum the most probable counts; summing the per-core posteriors instead overcounts sparse bands, whose training windows over-represent occupied cores.

\textit{Counting over $0.5$--$10$~mHz.}---The count is within $\pm1$ of the truth in $99.0\%$ of the 9330 cores at $\rth=8$ and $99.3\%$ at $\rth=10$, and the 95\% interval contains it in $99.1\%$ and $99.2\%$. Because the measured SNR of a binary scatters by about one, an oracle that knows every injected waveform and is ignorant only of the noise counts exactly in at most $90.7\%$ of the cores of $3.0$--$3.3$~mHz at $\rth=10$ ($77.6\%$ at $\rth=8$), and the flow reaches $90.8\%$ ($72.0\%$). Below $3.3$~mHz every flow comes within $1.1$ points of this limit at $\rth=10$ and $5.6$ at $\rth=8$ (Table~D1 of the data release): counting noise at the threshold, the GW analog of Eddington bias~\cite{Eddington1913}, is the only significant error. Above 6~mHz every binary lies far above threshold, the limit is $100\%$, and the flows count $93$--$98\%$ of the cores exactly, missing blends of bright, Doppler-broadened signals. In between, up to 14 bright binaries share a core and the flows undercount the most crowded cores, but their 95\% interval still contains the truth in $97$--$99\%$. In $3.3$--$6$~mHz the count is within $\pm1$ in $99.5\%$ of the cores with at most two binaries above $\rth=8$, $95.9\%$ with three to five and $67\%$ with six or more, which it undercounts by $0.7$ on average. Table~D1 of the data release gives, for each flow, the counts and the cores counted within $\pm1$ and exactly, with the information limit; counts in cells of $0.01$~mHz ($10^{-5}$~Hz) over the crowded band $3.3$--$6.0$~mHz and of $0.1$ or $0.5$~mHz elsewhere are in the data release.

\textit{Information limit.}---The oracle measures every binary's SNR as $x_i=\rho_i+n_i$ with $n_i\sim\mathcal{N}(0,1)$ and counts $x_i>c$; the cut $c$ is chosen within $\rth\pm2$ to maximize the fraction of exactly counted cores, averaged over noise realizations. No search can exceed it, since a search must also locate the binaries. For example, a binary of true SNR $10.3$ is measured below 10 with probability 38\% from the instrument noise alone.

\textit{Source counts.}---Because one evaluation covers every threshold, the flow also measures the source counts, a GW version of the $\log N$--$\log S$ test~\cite{Ryle1955}: on the 293 cores of $3.0$--$3.3$~mHz, its expected count gives the cumulative source count $N(>\rho)\propto\rho^{-\gamma}$ with $\gamma=3.39\pm0.22$, against the true $3.34\pm0.23$. A uniform population in Euclidean space gives $\gamma=3$ for any luminosity function, so the slope probes how the binaries are distributed in the Galaxy.

\subsection*{Stage 2: localization and sampling}

\textit{Localization.}---For each $K$ in the flow's range, an $F$-statistic~\cite{Jaranowski1998} matching pursuit~\cite{Mallat1993} adds binaries one at a time, as in iterative searches for Galactic binaries~\cite{CornishLarson2003}: with the four amplitude parameters of every binary solved analytically, the joint $F$-statistic is maximized over $(f_0,\dot f_0,\lambda,\sin\beta)$ of the new binary by differential evolution, and all binaries are then refined jointly by Levenberg--Marquardt. The Fisher matrix of the marginal likelihood at the joint mode defines, for each binary, coordinates obtained by a Cholesky factorization with unit Jacobian, which remove the strong correlations between $f_0$ and $\dot f_0$ and between $A$ and $\cos\iota$.

\textit{Sampling.}---Flow-assisted nested sampling with \texttt{nessai}~\cite{Skilling2006,Williams2021} samples Eq.~\eqref{eq:likelihood} under the threshold-truncated prior in a box of $\pm6\sigma$ around the mode in these coordinates, with 1000 live points, and the evidence inside the box enters Eq.~\eqref{eq:evidence}. Figure~\ref{fig:baseline} shows why the localization is needed: for four injected binaries (32 parameters), blind nested sampling over the full prior stalls at $\ln\Lmax=-1189$ after $1.2\times10^6$ likelihood calls in 67~min, far below the injected parameters ($-210.5$), whereas the localized sampler starts from the joint mode, found in 25~s, and converges in 33~min.

\begin{figure}[!htbp]
\centering
\includegraphics[width=\columnwidth]{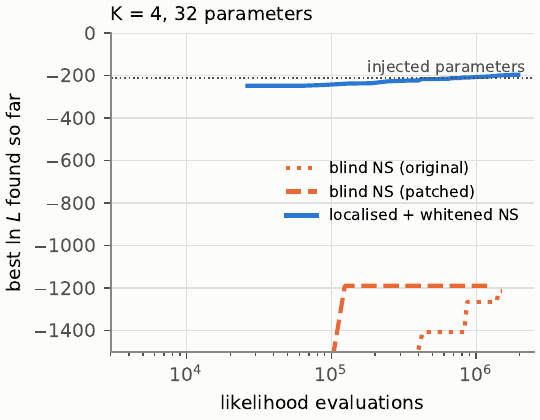}
\caption{Blind against localized nested sampling for four injected binaries (32 parameters), on the same data with the same likelihood: the highest log likelihood found so far against the number of likelihood calls. Two versions of a blind sampler over the full prior (dotted and dashed) stall more than 900 below the injected parameters (gray dotted line); the localized, Fisher-whitened sampler (blue) starts at the joint mode.}
\label{fig:baseline}
\end{figure}

\textit{Evidence ladder.}---The evidence ladder starts at the flow's mode and climbs or descends inside $[k_{\min},k_{\max}]$ while $\ln Z_K^{\rth}$ improves; Eq.~\eqref{eq:evidence} selects $K$, and the catalogue keeps the binaries of the middle core.

\textit{Degeneracy moves.}---After the localization, each binary in turn tries the ecliptic reflection $\sin\beta\to-\sin\beta$ and the sideband aliases $f_0\to f_0\pm1/\Tobs$, near-degenerate modes imparted by the orbital motion of the detector~\cite{Crowder2007,Littenberg2011}, alone and combined; the amplitudes are re-solved by the $F$-statistic and the fit refined, and the best move is kept if it raises $\ln\mathcal{L}$ by more than one. In the six injection windows this corrected one binary trapped in its mirror mode ($\Delta\ln\mathcal{L}=+4.4$, $\Delta\ln Z=+4.5$) and one trapped in a sideband alias ($\Delta\ln\mathcal{L}=+8.2$, $\Delta\ln Z=+11.0$). Rare configurations---two binaries at nearly the same sky position one sideband apart, next to a much brighter one---can still trap the localization.

\textit{Injection windows.}---Each of the six injection windows (Table~D2 of the data release) places three to five binaries in a core, adjacent ones 0.3--2 bins apart, with SNRs set to values between 10 and 28 and the other parameters drawn from their priors, in instrument noise only; the windows are 512 bins apart, so their segments do not overlap. Each window is sampled at $K_{\rm true}$ and $K_{\rm true}\pm1$ with $\rth=8$, and every posterior component is matched to an injected binary by the overlap of their whitened waveforms (Hungarian algorithm). The evidence selects the true number in every window, preferring it to one binary fewer by $\ln B=22$--$43$ and to one more by $\ln B=3.5$--$7.7$. Every binary is recovered with a waveform match of $0.959$--$0.997$, and 198 of the 208 injected values ($95.2\%$) lie inside the 95\% credible intervals. The median 68\% half-width of $f_0$ is $2.25$~nHz, with $\sigma_{f_0}\,\mathrm{SNR}\approx41$~nHz between the single-signal bounds for known ($17.5$~nHz) and free ($70$~nHz) $\dot f_0$; the 90\% sky areas span $3.8$--$56$~deg$^2$ (median $13$~deg$^2$) and fall roughly as $\mathrm{SNR}^{-2}$. In one year $\dot f_0$ is barely constrained at these frequencies, and $\phi_0$ and $\psi$ share a degenerate pair of modes, so their intervals are wide.

\subsection*{Counting first versus letting the evidence choose}

\textit{Counting and resolving.}---An ideally converged reversible-jump sampler draws $K$ from $P(K\,|\,d)\propto P(K)\,Z_K$ and, for a flat $P(K)$, settles on the $K$ of largest evidence. Fig.~D2(a) of the data release takes this $K$, over every $K$ of the evidence ladders, as the count of an ideal reversible-jump sampler: on 40 random windows it is exact in 63--78\% of the cores at $\rth=8$--$12$, against 78--95\% for the flow (Table~I of the Letter). Conversely, imposing a count on the sampler does not recover more binaries. On the 40 windows at $\rth=7$ (29 with every run available), fixing $K$ at the flow's mode or at the true number recovers 5 and 9 of the 12 binaries with SNR 7--8 and leaves 4 and 2 spurious components, against 12 and 1 when the evidence chooses $K$, although the flow's $K$ is correct in 22 of the 29 windows and the evidence's in 14. Counting and resolving are separate questions, which is why the flow sets the range of $K$ and the evidence chooses inside it.

\textit{The 40 windows.}---The 40 windows of Table~I of the Letter are middle cores drawn at random from the band, with non-overlapping segments. At each threshold Stage~2 is rerun with the prior truncated at $\rth$; the count of the ideal sampler is the number of middle-core components at the $K$ of largest evidence, the flow's count the mode of its posterior, and both are compared with the injected binaries above $\rth$ in the middle core, with standard errors over the 40 windows. The predicted biases use $\Occ=22.97$~nats and $\sigma=1.153$, measured on earlier windows; the procedure is detailed in the technical notes of the data release.

\subsection*{Theory of the ideal trans-dimensional count}

We derive the count that an ideally converged trans-dimensional sampler returns for threshold-defined sources, from its stationary distribution to a prediction of its accuracy, and illustrate each step with an example.

\textit{What an ideal trans-dimensional sampler returns.}---A reversible-jump sampler~\cite{Green1995} moves through the union of the parameter spaces of every $K$ and leaves invariant, by detailed balance, the joint posterior
\begin{equation}
p(K,\bm\theta_K\,|\,d)\propto P(K)\,\pi(\bm\theta_K\,|\,K)\,\mathcal{L}(d\,|\,\bm\theta_K,K).
\label{eq:rjtarget}
\end{equation}
Integrating over $\bm\theta_K$, the fraction of its samples with $K$ binaries converges to
\begin{equation}
\begin{gathered}
P(K\,|\,d)=\frac{P(K)\,Z_K}{\sum_{K'}P(K')\,Z_{K'}},\\
Z_K=\int\mathcal{L}(d\,|\,\bm\theta_K,K)\,\pi(\bm\theta_K\,|\,K)\,d\bm\theta_K.
\end{gathered}
\label{eq:pk}
\end{equation}
For a flat $P(K)$ two counts have posterior odds $Z_{K'}/Z_K=e^{\Delta\ln Z}$. Neighbouring counts differ by $3.5$--$43$~nats in the evidence ladders of the six injection windows, so an ideal sampler returns the $K$ of largest evidence with near certainty. We take this $K$ as its count (Table~I of the Letter); computing every $Z_K$ directly sidesteps the slow mixing of a real sampler.

\textit{The evidence of $K$ threshold-defined binaries.}---Defining sources by $\mathrm{SNR}>\rth$ truncates the prior of $K$ ordered binaries to
\begin{equation}
\pi_\rho(\bm\theta_K)=\pi(\bm\theta_K)\prod_{j=1}^{K}\frac{\Theta[\mathrm{SNR}(\bm\theta_j)-\rth]}{p_\rho},
\label{eq:truncprior}
\end{equation}
normalized by the prior probability $p_\rho$ that one binary exceeds $\rth$. The likelihood is negligible outside the localization box $B$, so $Z_K^{\rth}\simeq\int_B\mathcal{L}\,\pi_\rho\,d\bm\theta$. Inside $B$ we write $\pi=V_{\rm box}\,\pi_B$, with $\pi_B$ the prior normalized to the box and $V_{\rm box}$ its prior mass, and let $q_{\rm box}$ be the fraction of $\pi_B$ in which every binary exceeds $\rth$. Then $Z_K^{\rth}=(V_{\rm box}\,q_{\rm box}/p_\rho^{K})\,Z_{\rm box}$, where $Z_{\rm box}$ is the evidence that nested sampling computes under the truncated box prior $\pi_B\prod_j\Theta_j/q_{\rm box}$; this is Eq.~\eqref{eq:evidence}. The likelihood itself, Eq.~\eqref{eq:likelihood}, follows from the Gaussian likelihood $(2\pi\sigma^2)^{-N}e^{-Q/2\sigma^2}$ of the $2N$ real data values and the Jeffreys prior on $\sigma^2$,
\begin{equation}
\int_0^\infty\frac{e^{-Q/2\sigma^2}}{(2\pi\sigma^2)^{N}}\,\frac{d\sigma^2}{\sigma^2}=\frac{\Gamma(N)}{(2\pi)^N(Q/2)^N},
\label{eq:jeffreys}
\end{equation}
up to a constant that cancels in evidence ratios.

\textit{Inner product and likelihood gain.}---We write the noise-weighted inner product of two frequency series as
\begin{equation}
\langle a,b\rangle=\sum_{A,E}\sum_f\frac{\mathrm{Re}\,[a(f)\,b^*(f)]}{S_n(f)\,\Tobs/4},
\label{eq:inner}
\end{equation}
so that the optimal SNR of a waveform $h$ is $\rho=\langle h,h\rangle^{1/2}$, the residual power of Eq.~\eqref{eq:likelihood} is $Q=\langle r,r\rangle$ with $r=d-\sum_jh_j$, and noise of spectrum $\sigma^2S_n$ contributes $Q\approx2N\sigma^2$. At the maximum of Eq.~\eqref{eq:likelihood} the foreground level is $\hat\sigma^2=Q/2N$, so adding a $K$th binary raises the maximum marginal likelihood by
\begin{equation}
\Delta\ln\Lmax=N\ln\frac{\hat\sigma^2_{K-1}}{\hat\sigma^2_K}\simeq\frac{Q_{K-1}-Q_K}{2\hat\sigma^2}.
\label{eq:gain}
\end{equation}
Let $u$ be the waveform of the new binary at its best-fit parameters, normalized to $\langle u,u\rangle=1$, $r$ the residual of the other $K-1$ binaries, and $\hat\rho=\langle r,u\rangle$ its measured SNR. Fitting the binary with SNR $a$ leaves
\begin{equation}
Q(a)=\langle r-au,\,r-au\rangle=Q_{K-1}-2a\hat\rho+a^2.
\label{eq:quadratic}
\end{equation}
Without a threshold the minimum lies at $a=\hat\rho$, so $Q_{K-1}-Q_K=\hat\rho^{\,2}$ and
\begin{equation}
\Delta\ln\Lmax\simeq\frac{\hat\rho^{\,2}}{2\hat\sigma^2}.
\label{eq:freegain}
\end{equation}

\textit{Information cost.}---Near its maximum, the posterior of the new binary is close to a Gaussian with covariance $\Sigma$ in its eight parameters, and its prior is uniform, $\pi=1/V_{\rm prior}$, in the sampling coordinates. Laplace's approximation of the evidence integral $Z=\int\mathcal{L}\,\pi\,d\bm\theta$ then gives
\begin{equation}
\Delta\ln Z\simeq\Delta\ln\Lmax-\Occ,\qquad\Occ=\ln\frac{V_{\rm prior}}{(2\pi)^4|\Sigma|^{1/2}},
\label{eq:laplace}
\end{equation}
which is Eq.~(1) of the Letter: $\Occ$ is the logarithm of the factor by which the data shrink the parameter volume of the binary, i.e.\ the information they provide about it. For the six injection windows (posteriors in Figs.~D4--D9 of the data release), the posterior widths---$\pm2.3$~nHz in $f_0$ within a 96-bin prior region, $13$~deg$^2$ at 90\% over the whole sky, $\pm0.07$ in $\log_{10}A$ over 3.5 decades and $\pm0.15$ in $\cos\iota$---contribute about $6$, $9$, $3$ and $2$~nats, and the poorly constrained $\dot f_0$, $\phi_0$ and $\psi$ about one more, so $\Occ\approx21$~nats. Over all 1461 binaries added along our evidence ladders, $\Delta\ln Z$ tracks $\Delta\ln\Lmax$ (correlation $0.97$--$0.99$) with $\Occ\approx20$~nats (median $19.5$), i.e.\ 28 bits per binary [Fig.~D2(b) of the data release].

\textit{Confusion limit.}---By Eqs.~\eqref{eq:freegain} and \eqref{eq:laplace}, the evidence rises only if $\hat\rho^{\,2}/2\hat\sigma^2>\Occ$, i.e.\ for binaries above the Bayesian confusion limit $\rsc=\sqrt{2\Occ}\,\hat\sigma$.

\textit{Selection function.}---The threshold-truncated prior requires $a\ge\rth$ in Eq.~\eqref{eq:quadratic}. For $\hat\rho\ge\rth$ the fit is unchanged; for $\hat\rho<\rth$ the minimum lies on the boundary $a=\rth$, and
\begin{equation}
\Delta\ln\Lmax\simeq\frac{2\hat\rho\,\rth-\rth^2}{2\hat\sigma^2}=\frac{\hat\rho^{\,2}-(\rth-\hat\rho)^2}{2\hat\sigma^2},
\label{eq:boundarygain}
\end{equation}
so a binary just below the threshold, fitted at the boundary, loses only $(\rth-\hat\rho)^2/2\hat\sigma^2$ against the free fit. Treating $\Occ$ as constant, the evidence then admits binaries with measured SNR above
\begin{equation}
\rho^*(\rth)=\frac{\Occ\hat\sigma^2+\rth^2/2}{\rth}=\frac{\rth^2+\rsc^2}{2\rth}\qquad(\rth\ge\rsc),
\label{eq:selection}
\end{equation}
which lies below $\rth$ by $(\rth^2-\rsc^2)/2\rth$: above $\rsc$, the evidence also counts the binaries between $\rho^*$ and $\rth$. For $\rth<\rsc$, binaries with $\rth<\hat\rho<\rsc$ cannot pay their cost by Eq.~\eqref{eq:freegain}, the boundary is $\rho^*=\rsc$, and the evidence misses them. Evidence and source definition agree only at $\rth=\rsc$.

\textit{Predicted bias.}---The measured SNR of a binary of true SNR $\rho$ is $\hat\rho=\rho+\langle n,u\rangle$, where $\langle n,u\rangle$ is Gaussian with variance $\sigma^2$, so the evidence detects it with probability $\Phi[(\rho-\rho^*)/\sigma]$, with $\Phi$ the standard normal cumulative distribution. In a core whose injected binaries have true SNRs $\rho_i$, the expected count of the evidence alone minus the true count is therefore
\begin{equation}
b(\rth)=\sum_i\Phi\!\left[\frac{\rho_i-\rho^*(\rth)}{\sigma}\right]-\sum_i\Theta(\rho_i-\rth).
\label{eq:bias}
\end{equation}

\textit{Count distribution.}---The binaries of a core are detected independently, binary $i$ with probability $p_i=\Phi[(\rho_i-\rho^*)/\sigma]$, so the count of the ideal sampler follows a Poisson-binomial distribution,
\begin{equation}
P(K_{\rm ev}=k)=\sum_{|S|=k}\ \prod_{i\in S}p_i\prod_{i\notin S}(1-p_i),
\label{eq:countpmf}
\end{equation}
where $S$ runs over the subsets of the binaries in the core; we evaluate it by convolving their Bernoulli distributions. The probability of counting the core exactly is $P(K_{\rm ev}=K_{\rm true})$ with $K_{\rm true}=\sum_i\Theta(\rho_i-\rth)$, and its average over cores predicts the exact fraction of Table~I of the Letter, as Eq.~\eqref{eq:bias} predicts its bias. The model assumes resolved, independent binaries and constant $\Occ$ and $\sigma$.

\textit{Example 1: one binary near the threshold.}---At $\rth=10$, with $\Occ=22.97$~nats and $\sigma=1.153$ measured on earlier windows ($\rsc=7.81$, $\rho^*=8.05$), a binary of measured SNR $\hat\rho=7$, 8, 9, 10 and 11 raises the likelihood by $15.0$, $22.6$, $30.1$, $37.6$ and $45.5$~nats [Eq.~\eqref{eq:boundarygain} below $\rth$, Eq.~\eqref{eq:freegain} above], so $\Delta\ln Z=-7.9$, $-0.4$, $+7.1$, $+14.6$ and $+22.5$~nats. The ideal sampler therefore adds the binary of SNR 9, which the source definition excludes, with posterior odds of $1.2\times10^3$, and rejects that of SNR 7 with odds of $3.6\times10^{-4}$: this is the over-count above $\rsc$.

\textit{Example 2: four overlapping binaries.}---Without a threshold, a scale-free amplitude prior leaves a fifth, absent binary the probability $P(K=5)\approx0.28$ (Fig.~D1 of the data release): the prior volume of signals too faint to exclude dominates Eq.~\eqref{eq:pk}. Truncating the prior at $\rth=8$ gives $\ln(Z_4/Z_5)=+10.7$ and $\ln(Z_4/Z_3)=+31.8$, so $P(K=4\,|\,d)\simeq1-2\times10^{-5}$.

\textit{Example 3: an evidence ladder.}---In one of the 40 windows of Table~I of the Letter (the middle core starting at bin 94800, at $\rth=10$), the middle core holds a single binary, of true SNR $9.7$, so the true count is zero. The evidence ladder gives $\ln Z_K=-362.7$, $-298.5$, $-269.8$ and $-290.4$ for $K=0$--$3$. The first component fits a binary of SNR $14.8$ in an outer core, and the second the binary of SNR $9.7$, whose posterior SNR is pinned at $10.0$--$10.5$. From $K=1$ to $2$ the foreground level falls from $\hat\sigma^2=1.066$ to $0.877$, so Eq.~\eqref{eq:gain} gives $\Delta\ln\Lmax=256\ln(1.066/0.877)=50.0$~nats, within the $44$--$54$~nats of Eq.~\eqref{eq:boundarygain} for $\hat\rho\approx9.7$ and these values of $\hat\sigma^2$. The evidence rises by $28.7$~nats, so the binary costs $\Occ=21.3$~nats and is admitted; a third component gains only $15.2$~nats of likelihood and lowers the evidence by $20.6$~nats. The evidence thus selects $K=2$, with $P(K=2\,|\,d)=1-1.1\times10^{-9}$ by Eq.~\eqref{eq:pk}, and counts one binary in the middle core, where the source definition counts none, whereas the flow gives $P(K=0)=0.93$.

\textit{Example 4: forty windows.}---Table~D3 of the data release compares Eqs.~\eqref{eq:bias} and \eqref{eq:countpmf} with the measured count of the ideal sampler on the 40 windows of Table~I of the Letter. With $\Occ=22.97$~nats and $\sigma=1.153$ measured on earlier windows, the predicted fraction of exactly counted cores agrees with the measurement within $2.3$ points at $\rth=8$--$12$ ($78.9$, $69.8$, $65.6$ and $61.2\%$ against $77.5$, $67.5$, $67.5$ and $62.5\%$). At $\rth=7$, where nearly every binary lies between $\rth$ and $\rho^*$, the prediction is most sensitive to $\Occ$ and $\sigma$ and falls short ($70.9$ against $90.0\%$); the values measured on these runs, $\Occ=19.5$~nats and $\sigma=1.06$, give $89.6\%$ but overpredict the bias at $\rth=10$ and 12. No single pair of values fits every threshold, consistent with a foreground level that falls as a lower threshold admits more binaries into the model.

\subsection*{Joint posteriors}

The joint posteriors of the six injection windows, sampled at the true $K$, and of two dense windows of the $3.0$--$3.3$~mHz catalogue at the $K$ selected by the evidence are Figs.~D4--D11 of the data release; each shows the eight parameters of every binary with the injected values, the evidence differences to $K\pm1$ and the waveform match of each binary. Figure~\ref{fig:denseposterior} shows the largest joint fit, the 14 components of the dense segment of $4.2$--$4.6$~mHz (Dense segments, below).

\subsection*{Stage 2 in every band}

Outside $3.0$--$3.3$~mHz one window per band is inverted at the lowest flow threshold $\rho_8\ge8$ at which the flow's range ends at $K\le8$ in at least 95\% of the band's segments (16 at $4.2$--$7.0$~mHz), within $[k_{\min},\min(k_{\max},8)]$ (Fig.~D14 of the data release). Below 3~mHz and at $6$--$7$ and $8$--$10$~mHz every binary above $\rho_8$ is recovered, with matches $0.91$--$1.0$, and the middle core is left at or below the ideal residual, the data minus the true waveforms of every injected binary above $\rho_8$. Binaries are missed where they crowd: in $3.0$--$3.7$~mHz three of five and five of eight are recovered, the missed ones mostly in close groups, and in $4.2$--$6.0$~mHz, where the windows hold eight to eleven binaries above 16, the ladder, capped at eight, recovers three to seven. Of the 17 spurious components, 15 lie in the outer cores, which the catalogue drops, and 11 within 3.5 bins of a prior edge, most of them absorbing the leakage of a bright binary just outside it. Over the 744 catalogue ladders of $3.0$--$3.3$~mHz at $\rth=7$--$12$, every binary above $\rth$ is recovered in 92\%, 86\%, 73\%, 50\% and 32\% of the segments holding one to five, with median match $0.95$--$0.97$: Stage~2 is limited by how closely binaries crowd relative to their SNR rather than by their number alone. The results per band, three contiguous windows above 7~mHz and the capacity of the ladder are detailed in the technical notes of the data release.

\subsection*{Dense segments}

\textit{Iterating Stage 2.}---Where a segment holds more binaries than one fit can resolve, Stage~2 can count and resolve them in steps, brightest first, because the flows count above any threshold. The flow first counts the binaries above $\rth=16$, the evidence ladder resolves them, and their maximum-likelihood waveforms are subtracted; the residual is counted again at the lowest of the thresholds 16, 14, 12, 11 and 10 at which the flow's range ends at $K\le8$, resolved and subtracted, until the flow counts nothing above the catalogue threshold $\rth=10$; finally each block of binaries is refitted with the others subtracted. We test it on a segment of $4.2$--$4.6$~mHz whose flow count at $\rth=10$ is 16: its 96 bins hold 15 injected binaries above 10, one per six bins, six of them in the middle core (Fig.~\ref{fig:denseposterior}).

\textit{Result.}---Two passes, eight binaries above 16 and then six above 10, and the refit recover 12 of the 15 with matches $0.93$--$0.999$ and two spurious components (Fig.~\ref{fig:denseposterior}), and leave the middle core at $1.35$ times the noise level, against $1.13$ for the ideal residual and $1.05$ for the noise realization. One missed binary is fitted with match $0.79$, just below the cut; the other two lie in a group of four binaries within $1.8$ bins, between the two brightest (SNR 36 and 27), which are recovered. Localization alone, without nested sampling, recovers 9 of the 15 with five spurious components. In the other three segments of $4.2$--$4.6$~mHz with a flow count of 16, the same passes recover 7, 8 and 5 of their 16 binaries above 10, with 6, 4 and 8 spurious components.

\begin{figure*}[!htbp]
\centering
\includegraphics[width=\textwidth]{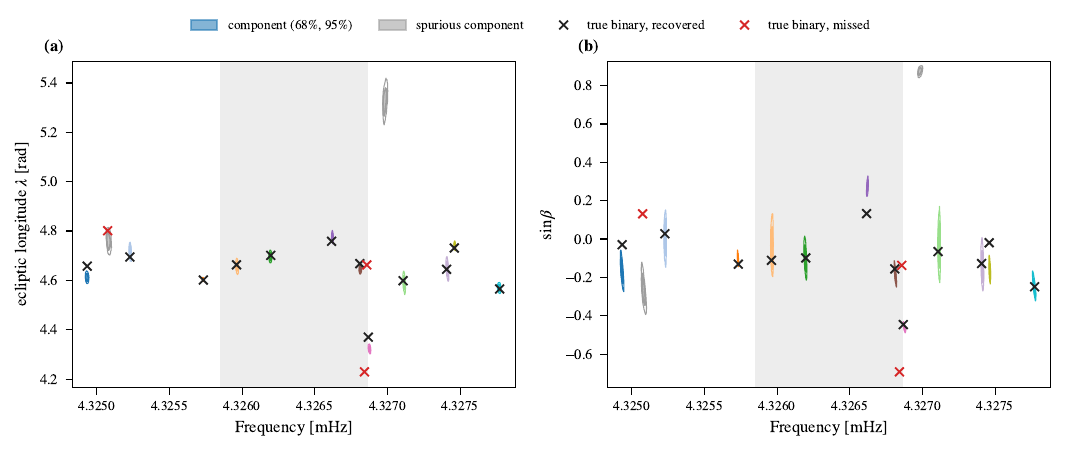}
\caption{Joint posterior of the dense segment of $4.2$--$4.6$~mHz after iterative Stage~2, the largest joint fit of this work: 68\% (filled) and 95\% contours of the 14 fitted components in frequency against (a) ecliptic longitude and (b) the sine of ecliptic latitude (spurious components gray; the frequency widths, 2--8~nHz, are narrower than the lines), with the 15 injected binaries above SNR~10 as crosses, black when recovered and red when missed; middle core shaded. The spectrum of the segment is Fig.~3(a) of the Letter.}
\label{fig:denseposterior}
\end{figure*}

\textit{The densest segment of every band.}---In each of the 13 bands we analysed the segment with the highest flow count at $\rth=10$ with the same passes (Fig.~3(b) of the Letter and Fig.~\ref{fig:bandmax}). The flows overcount in these segments, which hold 98 binaries above 10 in all; the passes recover 70: every one in the five bands whose segment holds two or three, eight of eight and six of seven at $6$--$8$~mHz, and 53--71\% in the crowded bands of $3.3$--$5.5$~mHz, whose segments hold 8 to 15. Of the 57 spurious components, 43 lie in the outer cores, which the catalogue drops.

\begin{figure*}[!htbp]
\centering
\includegraphics[width=\textwidth]{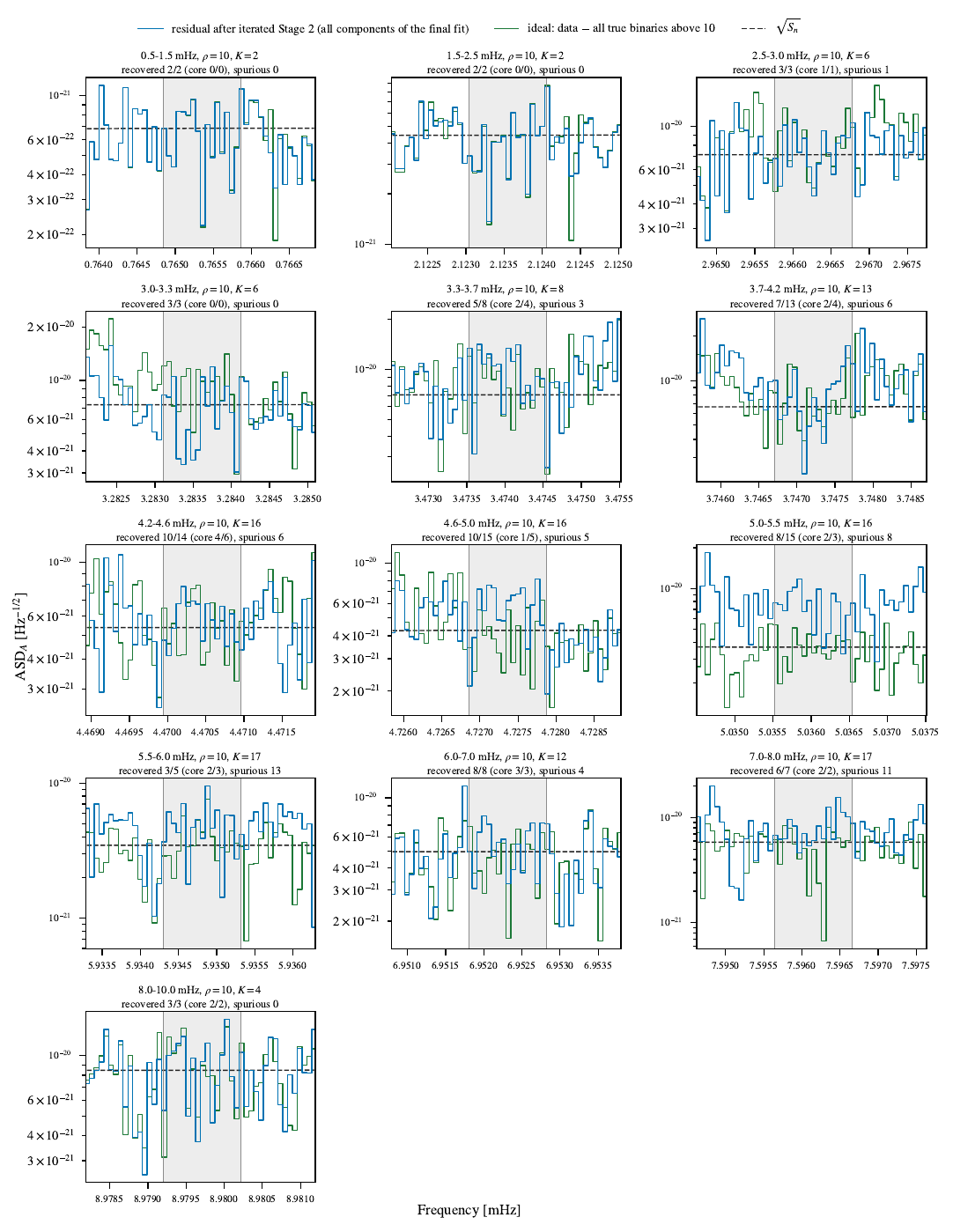}
\caption{Residual spectra of the densest segment of every band after iterative Stage~2, the full band of Fig.~3(b) of the Letter: ASD of channel $A$ over the prior region, averaged over pairs of bins (middle core shaded), of the residual after subtracting every component of the final fit (blue), the ideal residual, the data minus every injected binary above $\rth=10$ (green), and $\sqrt{S_n}$ (dashed). Panel titles give the number $K$ of fitted components, the binaries above 10 recovered over the true ones (middle core) and the spurious components. In the six crowded bands of $3.3$--$6.0$~mHz the joint fit of neighbouring segments lowers the residual further (Fig.~3(b) of the Letter; spectra in Fig.~D17 of the data release).}
\label{fig:bandmax}
\end{figure*}

\textit{Joint fit of neighbouring segments.}---The densest segments of the six crowded bands of $3.3$--$6.0$~mHz were then fitted jointly with their neighbours (Table~D4 and Fig.~D17 of the data release). Five overlapping segments, whose middle cores lie $-64$, $-32$, $0$, $32$ and $64$ bins from that of the densest segment, are the blocks of a Gibbs iteration: each block is fitted with Stage~2 on the data minus the middle-core components of the other four and keeps only the components of its own middle core, with every $K\le8$ from two below to two above the flow's most probable count of its residual sampled in full and the $K$ of largest evidence kept. Each of at most three sweeps updates the blocks at $0$ and $\pm64$ bins and then those at $\pm32$ (details in the technical notes of the data release). Over the six segments the joint fit recovers 60 of the 70 binaries above 10, against 43 after iterating, and 20 of the 25 in the middle cores, against 13; it fits 10 spurious components instead of 41, none of them at a prior edge, and lowers the middle-core residual in every band. At $4.2$--$4.6$, $4.6$--$5.0$ and $5.5$--$6.0$~mHz it recovers 13 of 14, 15 of 15 (the best recovered of them in Fig.~\ref{fig:jointcorner}) and 5 of 5 binaries, and the middle-core residual falls below the noise realization ($0.72$ against $0.98$, $0.68$ against $0.93$ and $1.18$ against $1.39$), and stays above the floor expected from a correct fit, which absorbs the noise its waveforms can describe ($0.69$, $0.62$ and $1.13$). At $3.3$--$3.7$ and $3.7$--$4.2$~mHz the residual falls even below the ideal one ($1.19$ against $1.55$ and $1.25$ against $1.38$), because the missed binaries are fitted as blends with fainter neighbours; at $5.0$--$5.5$~mHz the third sweep brings the middle core from $3.75$ to $1.12$ times the noise level, against an ideal $1.06$, recovering 12 of the 15 binaries. The 68\% and 95\% intervals cover 62\% and 83\% of the parameters of the recovered binaries, against 63\% and 81\% after iterating.

\begin{figure*}[!htbp]
\centering
\includegraphics[width=0.85\textwidth]{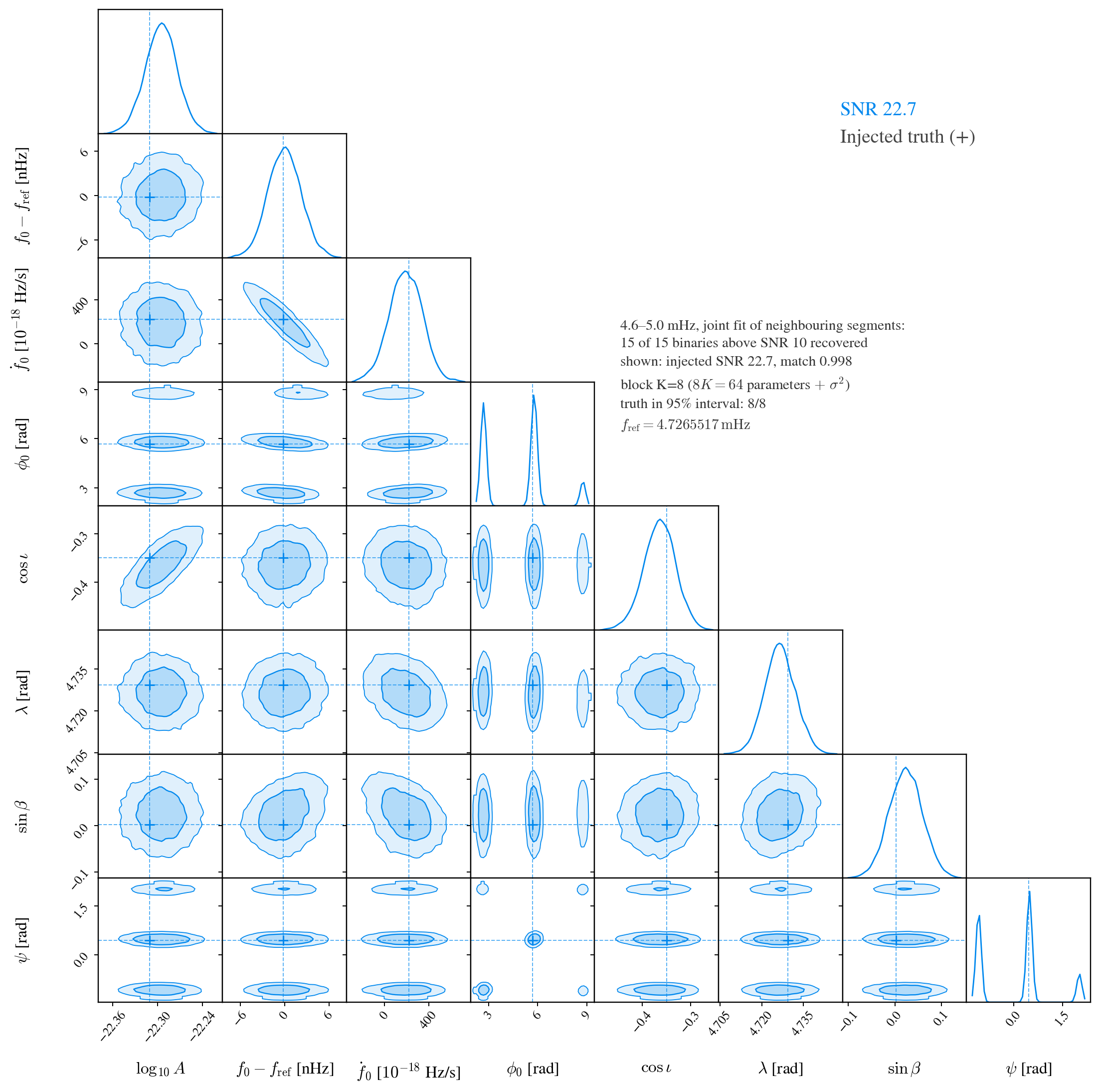}
\caption{Joint posterior of one of the 15 binaries above SNR~10 found by the joint fit of neighbouring segments at $4.6$--$5.0$~mHz, the best recovered: of the binaries whose eight injected values all lie inside the 95\% intervals, the one with the highest waveform match ($0.998$, injected SNR $22.7$), sampled together with the other seven components of its block. Contours enclose 68\% and 95\% of the posterior; crosses and dashed lines mark the injected values. Over all 15 binaries the matches are $0.819$--$0.999$, and 90 of the 120 injected values lie inside the 95\% intervals. $f_0$ and $\dot f_0$ are anticorrelated ($r=-0.94$) because $f_0$ refers to the start of the observation while the data fix the frequency at mid-observation, $f_0+\dot f_0\Tobs/2$; $\phi_0$ and $\psi$ show their usual degenerate modes.}
\label{fig:jointcorner}
\end{figure*}

\subsection*{Data release}

Figures~D1--D17 and Tables~D1--D4, cited above, are released with their captions at \url{https://github.com/aspirinliang/gbprl} (directory \texttt{data\_release}): model selection with and without the threshold (Fig.~D1); counting first against letting the evidence choose (D2); the catalogue of $3.0$--$3.3$~mHz (D3); the joint posteriors of the six injection windows and of two catalogue windows (D4--D11); the residual of $3.0$--$3.3$~mHz (D12, D13); Stage~2 in one window per band, in the dense segment, in the densest segment of every band and after the joint fit (D14--D17); counting by the 13 band flows (Table~D1), the injection windows (D2), the ideal trans-dimensional count (D3) and the joint fit (D4). Further tables are released as CSV files: binaries counted in frequency cells of $0.01$--$0.5$~mHz; Stage~2 in one window per band, its reruns at $K=9$ and its brightest-first passes; the densest segment of every band after iterated Stage~2; and the posterior median, 68\% half-width and injected value of every parameter of the six injection windows. Further details on Table~I of the Letter, on the catalogue of $3.0$--$3.3$~mHz and its residual, and on Stage~2 in every band and in the dense segments are in the technical notes, \texttt{technical\_notes.md}.

\fi

\end{document}